\documentclass[aps,reprint,nofootinbib,superscriptaddress]{revtex4-1}

\usepackage[utf8x]{inputenc}
\usepackage{amsmath}
\usepackage{amsfonts}
\usepackage{textcomp}
\usepackage{amssymb}
\usepackage{empheq}
\usepackage{esvect}
\usepackage{tikz}
\usepackage{subcaption}
\usetikzlibrary{calc}
\usepackage{xcolor}
\definecolor{lightpink}{RGB}{255, 230, 230}
\usepackage{hyperref}
\hypersetup{
  colorlinks   = true, 
  urlcolor     = blue, 
  linkcolor    = blue, 
  citecolor   = red 
}

\newcommand{\p}{\partial}
\newcommand{\f}{\frac}

\newcommand{\e}{\epsilon}
\newcommand{\be}{\beta}
\newcommand{\al}{\alpha}
\newcommand{\rot}{\operatorname{rot}}
\newcommand{\ve}{\boldsymbol}
\newcommand{\new}[1]{#1}
\begin{document}
\title{Non-radiating sources, dynamic anapole and Aharonov-Bohm effect}
\author{Nikita A. Nemkov}
\email{nnemkov@gmail.com}
\affiliation{Moscow Institute of Physics and Technology (MIPT), 141700 Dolgoprudny, Moscow region, Russia}
\affiliation{National University of Science and Technology (MISiS), The Laboratory of Superconducting metamaterials, 119049 Moscow, Russia}
\author{Alexey A. Basharin}
\email{alexey.basharin@gmail.com}
\affiliation{National University of Science and Technology (MISiS), The Laboratory of Superconducting metamaterials, 119049 Moscow, Russia}
\author{Vassily A. Fedotov}
\email{vaf@orc.soton.ac.uk}
\affiliation{Optoelectronics Research Centre, University of Southampton, Southampton SO17 1BJ, UK}

\begin{abstract}

\new{We show that for a particular choice of gauge the vector potential of any non-radiating source is spatially localized along with its electric and magnetic fields. Important on its own, this special property of non-radiating sources dramatically simplifies the analysis of their quantitative aspects, and enables the interpretation of non-radiating sources as distributions of the elementary dynamic anapoles. Using the developed approach we identify and discuss a possible scenario for observing the time-dependent version of the Aharonov-Bohm effect in such systems.} 
\end{abstract}

\maketitle

\section{Introduction}

This paper is concerned with properties of \emph{non-radiating} (NR) sources within the scope of the classical electrodynamics. We define an NR source as an oscillating charge-current configuration of a finite size, which does not generate any fields outside the volume it occupies. An alternative way of defining an NR source is to request that no energy is to be emitted into the far-field zone. However, as shown in \cite{devaney1973radiating}, this seemingly less restrictive definition also implies that the electromagnetic fields of the NR source are localized, i.e. they vanish outside the source volume. The interest in NR sources arose at the beginning of the twentieth century in the context of extended electron models, electromagnetic self-force and radiation reaction (see \cite{goedecke1964classically} and references therein). More recently, NR sources have become the subject of interest in relation to the inverse scattering problem of electrodynamics, i.e. reconstruction of sources from radiated fields (see \cite{friedlander1973inverse, devaney1973radiating, bleistein1977nonuniqueness, hoenders1997existence, marengo2000inverse, marengo2000nonradiating} for some representative works, and \cite{gbur2003nonradiating} for a review).

An example of a non-trivial yet simple NR source was theoretically proposed in the context of the so-called toroidal multipoles, the third independent family of dynamic multipoles that complement the conventional electric and magnetic ones (see for example \cite{afanasiev1998some}, \cite{costescu1987induced}). In particular, it was noted that the emissions of toroidal and electric dipoles have the same angular distribution and parity properties. Correspondingly, the electromagnetic fields radiated by coherently oscillating point toroidal and electric dipoles placed at the origin could be made to interfere destructively and disappear everywhere apart from the origin \cite{afanasiev1995electromagnetic}. This combination of interfering toroidal and electric dipoles forms a non-trivial point-like NR source, which is also known as the elementary \emph{dynamic anapole} (DA) \cite{dubovik2000material, raybould2015focused}. 

Despite its exotic appearance DA is anything but an abstract concept. First demonstrated experimentally in a specially designed microwave metamaterial, it was shown to play a key role in a new mechanism of electromagnetic transparency and scattering suppression \cite{fedotov2013resonant}. More recent works have confirmed the importance of DA also in the realms of plasmonics and nanophotonics, where dominant contributions of DAs were identified in the optical response of very simple types of dielectric and metallic nano-structures, such as discs and wires etc \cite{basharin2015dielectric, miroshnichenko2015nonradiating, liu2015toroidal, kim2015subwavelength, liu2015invisible, grinblat2016enhanced, xiang2016generic, tasolamprou2016toroidal, evlyukhin2016optical}.

\new{In this paper we show that for a particular choice of  gauge the vector potential of an NR source is localized, just as its electric and magnetic fields are. Exploiting the localization of potentials as the defining property of NR sources helps, for example, to find constraints on the actual current density in a relatively simple fashion, without using the heavy machinery of the multipole expansion \cite{radescu2002exact}. It allows one to consider NR sources as distributions of elementary DAs (in the same way as any radiating source can considered as a distribution of point charges), and therefore helps to build intuition about the internal structure of NR sources enabling the construction of explicit realizations.}

\new{Our approach provides a powerful alternative to that used by Devaney and Wolf in \cite{devaney1973radiating}, who first obtained the necessary and sufficient condition for an electromagnetic source to be non-radiating. It was formulated as a constraint on the Fourier components of oscillating current density. Since NR sources and their fields are, by definition, localized in space, the customary language of the Fourier modes, which are non-localized plane waves, may not always be a convenient choice. Working directly in the coordinate rather than momentum space (as it is done in the present work) should simplify the analysis and yield a clearer physical picture.}

We also conclude that NR sources provide a viable platform for observing the time-dependent Aharonov-Bohm effect. This idea had been originally proposed in \cite{afanasiev1995electromagnetic} but was met with skepticism by some authors, who argued that the dynamic version of the Aharonov-Bohm effect could not exist \cite{marengo2002nonradiating}. Using an explicit design of a finite-size NR source we show that some of the assumptions made in \cite{marengo2002nonradiating} may be relaxed, and that the key signature of the static Aharonov-Bohm effect will be present in the dynamic case.

\section{Elementary dynamic anapole}
\new{Before discussing general NR sources we describe the simplest example known as the elementary dynamic \textit{anapole} (DA), which is formed by collocated electric and toroidal point dipoles.}

A dynamic electric dipole $\ve{d}$ corresponds to the following spatial distributions of time-dependent charge and current density 

\begin{eqnarray}
\rho_{d}=-(\ve{d}\cdot\ve{\nabla})\,\delta(\ve{r}), \quad \ve{j}_d=\p_t\ve{d}\,\delta(\ve{r}), \label{el dip current}
\end{eqnarray}
while dynamic toroidal dipole $\ve{\tau}$ corresponds to
\begin{eqnarray}
\rho_{\tau}=0,\quad \ve{j}_{\tau}=c\rot^2{\ve{\tau}\delta(\ve{r})} \label{tor dip current}
\end{eqnarray}

If $\ve{d} = c^{-1}\p_t\ve{\tau}$ then these two elementary sources are known to produce exactly the same electric and magnetic fields everywhere except for $\ve{r}=0$ \cite{afanasiev1995electromagnetic}. They also give rise to electromagnetic potentials, which are gauge equivalent beyond $\ve{r}=0$.
In principle, this allows one to create a dynamic source that does not radiate. Without loss of generality we assume harmonic time-dependence with frequency $\omega$, and hence we replace $\ve{d}\to\ve{d}e^{-i\omega t}, \ve{\tau}\to\ve{\tau}e^{-i\omega t}$ with $\ve{d}$ and $\ve{\tau}$ now being constant vectors. Electric and magnetic fields of the two dipoles, which are placed at the same point, will interfere destructively provided that 

\begin{eqnarray}
\ve{d}=-ik\ve{\tau}, \label{d to tau relation}
\end{eqnarray}
with $k=\omega/c$. It is this configuration that yields the elementary dynamic anapole. Below, we will characterize elementary DA by its toroidal dipole moment $\ve{\tau}$ keeping in mind that it is always accompanied by an electric dipole moment \eqref{d to tau relation}. There is a gauge choice for which the potentials of the elementary DA become
\begin{align}
\phi_{DA}&=\phi_d+\phi_{\tau}=0 \label{anapole scalar potential}\\
 \ve{A}_{DA}&=\ve{A}_d+\ve{A}_{\tau}=e^{-i\omega t}4\pi\ve{\tau}\delta(\ve{r}) \label{anapole potential}
\end{align}

Electric and magnetic fields of DA can be obtained from the usual relations

\begin{eqnarray}
\ve{E}=-\ve{\nabla}{\phi}-\p_t\ve{A},\quad \ve{H}=\rot\ve{A}
\end{eqnarray}
which, taking into account \eqref{anapole scalar potential} and \eqref{anapole potential}, give
\begin{align}
\ve{E}_{DA}&=e^{-i\omega t}4\pi i k\ve{\tau}\delta(\ve{r})\label{anapole electric field}\\
\ve{H}_{DA}&=e^{-i\omega t}4\pi  \rot\ve{\tau}\delta(\ve{r})\label{anapole magnetic field}
\end{align}

By substituting \eqref{anapole electric field} and \eqref{anapole magnetic field} into Maxwell's equations one can easily verify that these fields indeed correspond to a combination of the electric and toroidal dipole currents given by \eqref{el dip current} and \eqref{tor dip current}.

\begin{figure}
\begin{subfigure}{0.5\textwidth}
\centering
\includegraphics[width=0.5\linewidth]{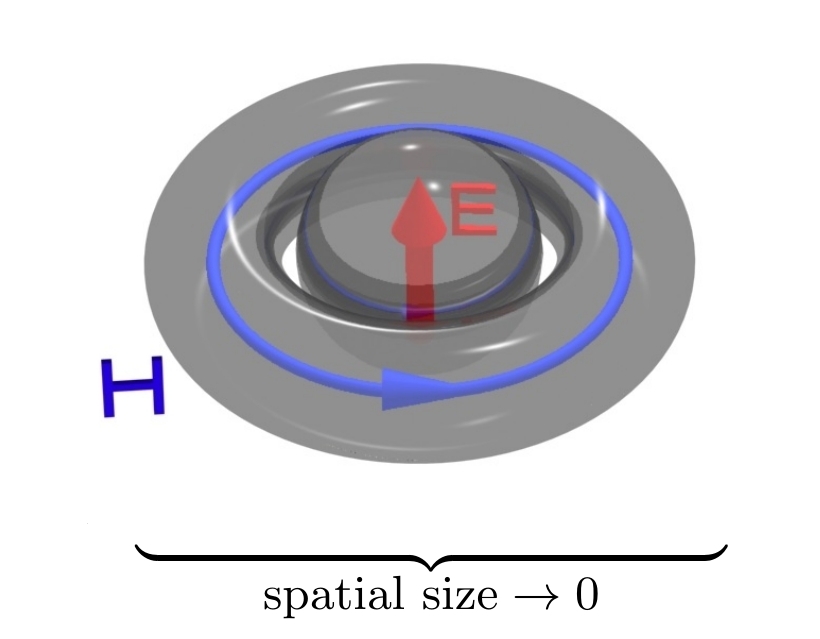}
\end{subfigure}
\caption{An artistic impression of the elementary dynamic anapole. The anapole is presented in terms of its electric and magnetic fields, and the volumes they occupy. In an elementary dynamic anapole the magnetic field forms an infinitesimal loop, which encirles the electric field confined to a point.}
\label{elementary anapole}
\end{figure}

It will prove useful to visualize electric and magnetic fields of the elementary DA (see fig. \ref{elementary anapole}). Electric field points in the direction of the toroidal moment $\ve{\tau}$ while the magnetic field is represented by an infinitesimal loop in the plane orthogonal to the electric field.

\section{General non-radiating sources \label{general sources}}

\new{By definition, an NR source does not produce electric or magnetic fields outside the volume it occupies. We will now show that in a certain gauge the vector potential generated by an NR source is also zero outside the volume of the source.
}

Indeed, any electromagnetic field can be described by vector and scalar potentials $\phi, \ve{A}$. Due to gauge freedom the scalar potential can be always chosen vanishing $\phi=0$ (the Weyl gauge). Without loss of generality we restrict analysis to the vector potential with harmonic time behavior $\ve{A}(t,\ve{r})=e^{-i\omega t}\ve{A}(\ve{r})$. \new{Clearly, these two conditions specify the chosen gauge uniquely.\footnote{\new{The Weyl gauge is incomplete since it leaves residual gauge transformations $\ve{A}\to \ve{A}+\nabla{\chi}$ with arbitrary time-independent function $\chi$. The requirement that $\ve{A}$ must be harmonic in time eliminates this freedom.}} Now, in this gauge the electric field is given by the time-derivative of the vector potential $\ve{E}=-c^{-1}\p_t \ve{A}=ik\ve{A}$. Hereby, if the vector potential is non-zero at some point in space so is the electric field. Therefore, the vector potential of any NR source must vanish everywhere where the electric field does, i.e. outside the volume that the source occupies.\footnote{\new{This argument of course fails in the static case $k=0$ where non-trivial potentials can exist in the absence of fields.}} The latter suggests that for time-dependent NR sources the condition $\phi=0$ is a natural gauge fixing. It is in some sense the most economic gauge: the vector potential is only present where the electric or magnetic fields are non-vanishing. We will use the Weyl gauge throughout the paper, often without mentioning it explicitly.}

\new{In paper \cite{devaney1973radiating} the necessary and sufficient condition for a source to be non-radiating was formulated in terms of the Fourier components of the charge-current density. We prove in appendix \ref{nr criteria} that both formulations are in fact equivalent. Nevertheless, characterizing NR sources by their potentials can be advantageous from several standpoints. First of all, one is free to choose arbitrary localized vector potential and then find the corresponding current density describing the NR source at hand using Maxwell's equations. The latter is much easier than describing the NR source directly in terms of the charge-current density, which must satisfy non-trivial (and non-local) conditions derived by Devaney and Wolf in \cite{devaney1973radiating}. This particular advantage of our approach is clearly illustrated in the previous section: starting with the simplest possible form of the localized vector potential \eqref{anapole potential} one discovers the elementary DA, a quite non-trivial configuration of currents. Instead of specifying an NR source in terms of localized potentials one may also attempt to do the same using localized electric and magnetic fields. That, however, will be typically a more involving task since electric and magnetic fields must satisfy Maxwell's equations and therefore cannot be chosen arbitrarily. In contrast, the electromagnetic potentials are not constrained and in this sense represent independent degrees of freedom.} 
	
\new{Viewing NR sources in terms of potentials also provides some intuition about their properties and allows to construct explicit examples, as we will show in the next section.}

\new{Note that the vector potential of the elementary DA is proportional to the $\delta$-function \eqref{anapole potential}. Hence, the potential of the elementary DA may serve as a building block out of which arbitrary potential can be composed.
} Indeed, consider three DAs with their unit moments directed along Cartesian co-ordinate axes

\begin{eqnarray}
\ve{A}^\al_{DA}=e^{-i\omega t}\,\widehat{\ve{r}}^\al\delta(\ve{r})\label{basis anapoles}
\end{eqnarray}
Vector potential of an electromagnetic field can then be decomposed  in co-ordinate basis as $\ve{A}(t,\ve{r})=\sum_{\al=1}^3\widehat{\ve{r}}^\al A^\al(t,\ve{r})$. Correspondingly, for any vector potential $\ve{A}$ holds

\begin{multline}
\ve{A}(t,\ve{r})=\sum_{\al=1}^3\widehat{\ve{r}}^\al \int d\ve{r}'\, A^\al(t,\ve{r}')\delta(\ve{r}-\ve{r}')=\\\sum_{\al=1}^3\int d\ve{r}'\, A^\al(\ve{r}')\ve{A}^\al_{DA}(t,\ve{r}-\ve{r}')\label{potential from anapoles}
\end{multline}
This expression represents arbitrary vector potential as a superposition of the vector potentials due to the elementary DAs. Since this conclusion might seem counter-intuitive we make some clarifications in appendix \ref{clarifications}. 

\new{In equation \eqref{potential from anapoles} the integration effectively runs over the domain where the vector potential is non-vanishing. As shown above, for NR sources this domain has a finite volume. The latter implies that the corresponding charge-current density is composed out of the elementary DA densities
\begin{align}
\rho(t,\ve{r})=\sum_{\al=1}^3\int d\ve{r}'\, A^\al(\ve{r}')\rho^\al_{DA}(t,\ve{r}-\ve{r}')\\
\ve{j}(t,\ve{r})=\sum_{\al=1}^3\int d\ve{r}'\, A^\al(\ve{r}')\ve{j}^\al_{DA}(t,\ve{r}-\ve{r}')\label{densities from anapoles}
\end{align}
where 
\begin{align}
\rho^\al_{DA} &=e^{-i\omega t} \f{ik}{4\pi}(\widehat{\ve{r}}^\al\cdot\ve{\nabla})\, \delta(\ve{r})\\
\ve{j}^\al_{DA} &=e^{-i\omega t}\f{c}{4\pi}\left( \rot^2 \widehat{\ve{r}}^\al\delta(\ve{r})-k^2 \widehat{\ve{r}}^\al \delta(\ve{r})\right) \label{anapole current}
\end{align}
}

\new{Accepting that the vector potential of an NR source is localized also allowed us to check the validity of 
\eqref{d to tau relation} for spatially extended NR sources. Our analysis does not rely on the multipole expansion but merely uses the definitions of $\ve{D}$ and $\ve{T}$
\begin{eqnarray}
D_\al=-\f1{i\omega}\int d\ve{r}\, j_\al \label{D def}\\
T_\al=\f1{10c}\int d\ve{r}\, (r_\al r_\be-2r^2\delta_{\al\be})j_\be \label{T def}
\end{eqnarray}
We now show this assuming only the localization property. In the Weyl gauge, the current density is related to the vector potential as follows (in tensor notation)
\begin{eqnarray}
j_\al(\ve{r})=\f{c}{4\pi}\left(-(k^2+\Delta)\delta_{\al\be}+\nabla_\al\nabla_\be\right)A_\be(\ve{r}) \label{current via vector potential}
\end{eqnarray} 
Let us substitute \eqref{current via vector potential} to \eqref{D def}
\begin{multline}
D_\al=-\f{1}{4\pi ik}\int d\ve{r}\, \left(-(k^2+\Delta)\delta_{\al\be}+\nabla_\al\nabla_\be\right)A_\be=\\\f{-ik}{4\pi}\int d\ve{r}\, A_\al+\f{1}{4\pi ik}\int d\ve{r}\, \left(\delta_{\al\be}\Delta-\nabla_\al\nabla_\be\right)A_\be
\end{multline}
The last integral can be reduced to a surface integral by the Gauss theorem. Since the vector potential is localized the surface integral vanishes and one gets
\begin{eqnarray}
D_\al=\f{-ik}{4\pi}\int d\ve{r}\, A_\al
\end{eqnarray}
Similarly, substituting \eqref{current via vector potential} to \eqref{T def}, integrating by parts twice, and disregarding the boundary terms one arrives at
\begin{eqnarray}
T_{\al}=\f{1}{4\pi}\int d\ve{r}\, A_\al-\f{k^2}{40\pi}\int(r_\al r_\be -2r^2\delta_{\al\be})A_{\be} \nonumber\\\label{T res}
\end{eqnarray}
The last contribution can be estimated as 
\begin{eqnarray}
\frac{k^2\int(r_\al r_\be -2r^2\delta_{\al\be})A_{\be}}{\int d\ve{r}\, A_\al}=O(a^2k^2)
\end{eqnarray}
where $a$ is the spatial extent of the source. Hence, we obtain relation 
\begin{eqnarray}
\ve{D}=-ik\ve{T}(1+O(a^2k^2))\label{D to T general}
\end{eqnarray}
Although somewhat surprising, this result fully agrees with \cite{radescu2002exact}.
In general, relation \eqref{D to T general} cannot be satisfied exactly (unless $\ve{D}=0$), because $\ve{T}$, being a higher order multipole moment, depends on the origin of the multipole
expansion, and so the high-order correction $O(a^2 k^2)$ in \eqref{D to T general} arises as the result of this uncertainty}.

\section{Example of a spatially extended non-radiating source \label{extended anapole}}

\begin{figure}
\begin{center}
\begin{tikzpicture}

\draw[dashed] (0,0) ellipse (1.8 and 1.3);
\pgfmathsetmacro{\centx}{0}  
\pgfmathsetmacro{\centy}{0}
\draw[line width =1, blue, ->] (\centx,\centy) ellipse (0.3 and 0.2);
\draw[line width = 1, blue, ->] (\centx-0.3,\centy) --++ (0, -0.05); 
\draw[line width = 1, blue, ->] (\centx+0.3,\centy) --++ (0, +0.05);

\draw[line width = 1, red, ->] (\centx, \centy) --++ (0, 0.35);

\def\a{1.5}
\def\b{1}
\foreach \r in {0.5, 1}{
 \pgfmathsetmacro{\start}{30/\r}
 \pgfmathsetmacro{\step}{60/\r}
 \foreach \an in {\start,\step,...,360}{
  \pgfmathsetmacro{\centx}{\a*\r*cos(\an)}
  \pgfmathsetmacro{\centy}{\b*\r*sin(\an)}
  
  \draw[line width = 1, blue, ->] (\centx,\centy) ellipse (0.3 and 0.2);
  \draw[line width = 1, blue, ->] (\centx-0.3,\centy) --++ (0, -0.05); 
  \draw[line width = 1, blue, ->] (\centx+0.3,\centy) --++ (0, +0.05);
  \draw[line width = 1, red, ->] (\centx, \centy) --++ (0, 0.35);
  }
 }
\node at (2.5,0) {$\Rightarrow$};
 
\filldraw[color=lightpink, line width = 1.2] (5,0) ellipse (1.7 and 1.2);
\draw[dashed] (5,0) ellipse (1.8 and 1.3);
\draw[blue, line width = 1.2] (5,0) ellipse (1.7 and 1.2);
\draw[blue, line width = 2,->] (5-1.7,0) --++ (0,-0.01);
\draw[blue, line width = 2,->] (5+1.7,0) --++ (0,+0.01);
\pgfmathsetmacro{\centx}{5}  
\pgfmathsetmacro{\centy}{0}
\draw[line width = 1, red, ->] (\centx, \centy) --++ (0, 0.35);

\def\a{1.5}
\def\b{1}
\foreach \r in {0.5, 1}{
 \pgfmathsetmacro{\start}{30/\r}
 \pgfmathsetmacro{\step}{60/\r}
 \foreach \an in {\start,\step,...,360}{
  \pgfmathsetmacro{\centx}{\a*\r*cos(\an)+5}
  \pgfmathsetmacro{\centy}{\b*\r*sin(\an)}

  \draw[line width = 1, red, ->] (\centx, \centy) --++ (0, 0.35);
  }
 } 
 
\node at (0,-1.8) {(a)};
\node at (5,-1.8) {(b)};

\end{tikzpicture}
\end{center}
\caption{A schematic of an extended non-radiating source -- a disk uniformly filled with anapoles.}
\label{disk anapole}
\end{figure}
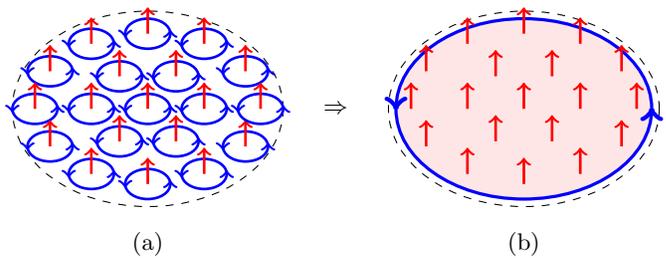

As a particular example let us consider a flat disk $D$ of radius $R$ uniformly filled with elementary DAs of surface density $\sigma$ (see fig. \ref{disk anapole}b) so that $\Delta\ve{\tau}=\ve{n}\sigma \Delta S$ is the toroidal dipole moment gathered in area $\Delta S$ with normal vector $\ve{n}$. The vector potential of this disk is a superposition of the potentials for the constituent DAs

\begin{multline}
\ve{A}(t,\ve{r})=\sum_{\al=1}^3\iint_{D} d^2s\,\sigma \,n^\al\ve{A}^\al_{DA}(t, \ve{r}-\ve{r}_s)=\\e^{-i\omega t}\sigma \ve{n}\iint_{D} d^2s\,\delta(\ve{r}-\ve{r}_s), \label{disk potential}
\end{multline}  
where $\ve{r}_s$ are position vectors of the points at the disk $D$.

Outside disk $D$ the electric and magnetic fields vanish. It is instructive to see how they are distributed within the disk. Electric field is computed as the time-derivative of the vector potential

\begin{eqnarray}
\ve{E}(t,\ve{r})=e^{-i\omega t}i\omega\sigma\ve{n}\iint_{D} d^2s\,\delta( \ve{r}-\ve{r}_s) \label{disk electric field}
\end{eqnarray}

It is homogeneous and directed along the normal vector $\ve{n}$. Magnetic field is given by

\begin{multline}
\ve{H}(t,\ve{r})=e^{-i\omega t}\sigma \iint_D d^2s\, \rot \ve{n}\,\delta(\ve{r}-\ve{r}_s)=\\e^{-i\omega t}\sigma \oint_C d\ve{l}\,\delta(\ve{r}-\ve{r}_l) \label{disk magnetic field}
\end{multline}

Stokes' theorem was used to rewrite surface integral as the integral over circle $C$ which is the boundary of disk $D$ and consists of points $\ve{r}_l$. We see that the magnetic field exists only at the boundary, and is constant in magnitude and oriented along the tangent line.

These results have a simple geometric explanation. If each constituent DA is visualized according to fig. \ref{elementary anapole}, then the non-radiating disk can be depicted as shown in fig. \ref{disk anapole}a. The electric fields of the adjacent anapoles do not interfere and result in a uniform total electric field. The magnetic fields of the adjacent anapoles are oriented oppositely and cancel each other, so the net magnetic field is zero everywhere within the area of the disk. The resulting field configurations are shown in fig. \ref{disk anapole}b, and are indeed described by expressions \eqref{disk electric field} and \eqref{disk magnetic field}, yielding an extended NR source. Note that the electric and magnetic fields, as well as the magnetic flux carried by the boundary magnetic line 

\begin{eqnarray}
\Phi=e^{-i\omega t}\sigma \label{anapole magnetic flux},
\end{eqnarray}
depend on the density $\sigma$ but not on the disk radius $R$.	

Fig. \ref{disk anapole}b can be regarded as non-radiating generalization of the field configuration of an ordinary static magnetic solenoid.
In the static case (frequency $\omega=0$) the electric field within the disk vanishes and only the boundary magnetic line with a constant flux remains. In the dynamic case, however, such a field configuration is supplemented by the electric field but only in the region encircled by the magnetic field.

Similarly to fig. \ref{disk anapole}, it is straightforward to visualize field configurations of more general NR sources. Indeed, formulas \eqref{disk potential}, \eqref{disk electric field}, \eqref{disk magnetic field} are valid for a flat layer $D$ with boundary $C$ of any shape (not necessarily a disk). Consequently, any three-dimensional domain filled with elementary anapoles homogeneously distributed over its volume V can be rendered as a stack of flat layers each of which is treated as above. It takes all but a small step to conclude that such a domain will feature homogeneous electric field in its volume and magnetic field confined to its boundary. Non-uniform electric field and non-vanishing magnetic field in the bulk can then be achieved by allowing anapole density and orientation to vary. Such more complex configurations can be considered as an overlap of homogeneous extended NR sources.

\section{Time-dependent Aharonov-Bohm effect \label{AB effect}}

We now turn to the discussion of the prospects which dynamic NR sources open in connection with the Aharonov-Bohm (AB) effect. The AB effect rests on the observation that in the quantum theory particles can be affected by electromagnetic interaction even if they do not contact electric or magnetic fields directly. 

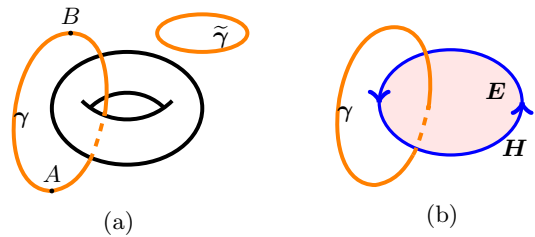
\begin{figure}
\centering
\begin{subfigure}{0.24\textwidth}
\centering
\begin{tikzpicture}[scale=0.5]

\draw[line width = 1.5] (0,0) ellipse (2 and 1.5);
\draw[line width = 1.5] (-1.2,.2) to [out=-45, in=-135] (1.2,.2);
\draw[line width = 1.5] (-1,0) to [out=45, in=135] (1,0);

\draw[orange, line width = 1.5] (-0.6,-0.2) to [out=80, in=0] (-1.5,2) to [out=180,in=180] (-2,-2.2) to [out=10,in=-120] (-1,-1.4);
\draw[orange,dashed,line width=1.5] (-1,-1.4) to [out=60,in=80] (-0.6,-0.2);

\filldraw (-1.5,2) circle (0.05);
\node[above] at (-1.5,2) {$B$};
\filldraw (-2,-2.2) circle (0.05);
\node[above] at (-2,-2.2) {$A$};
\node at (-2.8,-.3) {$\boldsymbol{\gamma}$};

\draw[orange,line width=1.5] (2,2) ellipse (1.2 and 0.5);
\node at (2.5,1.9) {$\boldsymbol{\widetilde{\gamma}}$};

\end{tikzpicture}
\caption{}
\label{ab toroid}
\end{subfigure}%
\begin{subfigure}{0.24\textwidth}
\centering
\begin{tikzpicture}[scale=0.5]
\filldraw[color=lightpink, line width = 1.2] (0,0) ellipse (1.9 and 1.4);
\draw[blue, line width = 1.2] (0,0) ellipse (1.9 and 1.4);
\draw[blue, line width = 2,->] (-1.9,0) --++ (0,-0.01);
\draw[blue, line width = 2,->] (+1.9,0) --++ (0,+0.01);

\draw[orange, line width = 1.5] (-0.6,-0.2) to [out=80, in=0] (-1.5,2) to [out=180,in=180] (-2,-2.2) to [out=10,in=-120] (-1,-1.4);
\draw[orange,dashed,line width=1.5] (-1,-1.4) to [out=60,in=80] (-0.6,-0.2);

\draw[white,line width=1.5] (2,2) ellipse (1.2 and 0.5);

\node at (1.2, 0.3) {$\ve{E}$};
\node at (1.7, -1.2) {$\ve{H}$};
\node at (-2.8,-.3) {$\boldsymbol{\gamma}$};
\end{tikzpicture}
\caption{}
\label{disk with contours} 
\end{subfigure}
\caption{Contours in the Aharonov-Bohm effect: (a) static case (b) dynamic case.}
\end{figure}

The most celebrated example of a system supporting the AB effect is a solenoid bent into a torus with a constant magnetic flux, see fig. \ref{ab toroid}. Magnetic field is only present inside the torus while electric field is absent everywhere. Probability amplitudes for a particle of charge $e$ to travel from point $A$ to $B$ along two paths, one of which lies inside and the other outside the torus hole, will have additional relative phase shift due to the vector potential 
\begin{eqnarray}
\delta \phi =e/\hbar c \oint_\gamma \ve{A}\,d\ve{r} \label{phase shift}
\end{eqnarray}
where the integral is taken along the contour $\gamma$ winding on the torus. By Stokes' theorem this integral is proportional to the magnetic flux $\Phi$ inside the torus $\delta \phi=e\Phi/\hbar c$. This phase shift has physically measurable consequences which were confirmed in many works, see e.g. \cite{peshkin1989aharonov}.

It is natural to attempt to generalize the AB effect, extending its reach towards the time-dependent case. In the context of the NR sources this question was previously addressed in \cite{marengo2002nonradiating, afanasiev1996interaction}. The most clear version of the effect would imply

\noindent (i) Presence of some volume $V$ inside which electric and/or magnetic field is non-zero (and time-dependent) but outside which both of them are absent. 
\newline (ii) Non-vanishing time-dependent phase shifts for some paths which lie outside $V$.

Requirement (ii) can be alternatively formulated as non-triviality of the electromagnetic potential outside $V$. Despite the fact that the AB effect is an essentially quantum phenomenon, our focus is on the classical electromagnetic fields and potentials. Thus, we are referring to phase shifts for brevity, as we mainly discuss the properties of classical fields.

As claimed in section \eqref{general sources} it is not possible to radiate the vector potential without also radiating the electric field, hence conditions (i, ii) are not possible to satisfy simultaneously. This leads to an immediate conclusion that the time-dependent version of the AB effect is simply not possible, at least not in its original form. However, such sharp contrast with the static case calls for an explanation.

One of the reasons behind discontinuity between static and dynamic situations originates from the fact that different contours must be considered. As mentioned earlier, the configuration in fig. \ref{disk anapole} serves as a dynamic non-radiating counterpart of a static toroidal solenoid\footnote{Instead of infinitesimally thin solenoid one is free to consider its realistic three-dimensional prototype. This has no effect on our conclusions but unnecessarily obstructs the computations.}. In the dynamic case the absence of radiation implies that the electric field is localized within the hole of the torus. Correspondingly, any contour penetrating the hole (as considered in the static case) will cross the region of non-zero electric field and therefore become ineligible in the context of the original effect. Other contours, which do not cross the field lines (similar to $\widetilde{\gamma}$ shown in fig. \ref{ab toroid}), will produce phase shifts neither in the static nor dynamic cases. This is because these contours do not encircle the region of non-zero magnetic field and therefore the integral \eqref{phase shift} has to vanish in accordance with Stokes' theorem. 

Let us consider a non-trivial contour $\gamma$ in the dynamic case, see fig \ref{disk with contours} (electric field arrows are suppressed). Assume that the particles in the Aharonov-Bohm experiment travel fast enough so that the fields do not change much during their flight. Then, we could use the same formula for the phase shift as in the static case \eqref{phase shift} provided that the vector potential is taken at the appropriate instant. \new{Applying Stokes' theorem to integral \eqref{phase shift} in the time-dependent case yields
\begin{eqnarray}
\delta\phi(t)=\f{e\Phi(t)}{\hbar c}=e^{-i\omega t}\f{e\sigma}{\hbar c} \label{time phase shift}
\end{eqnarray}
This phase shift is proportional to the time-dependent magnetic flux \eqref{anapole magnetic flux}.} Therefore, one can maintain the same interpretation for this phase shift as in the static case, i.e. conclude that it is solely due to the magnetic flux in the excluded region.

This observation is also interesting for the following reason. The naive way of producing a time-dependent phase shift would be to take an ordinary wire solenoid and vary the magnetic field with time, for example, by varying electric current in the windings. However, as noticed in paper \cite{Singleton:2013qva}, this trick would not work as the phase shift in fact remains constant. This is due to the fact that these time-varying currents are bound to produce electromagnetic field outside the solenoid. Its contribution to the phase shift appears to cancel exactly the time-dependent part resulting from the magnetic flux inside the torus. This again highlights the peculiarity of NR sources where the time-dependent phase shift arises naturally.

One might raise a natural objection to calling the described thought experiment as time-dependent AB effect. After all, the local impact of the electric field is still present. It is easy to see though, that this impact cannot account for the phase shift \eqref{time phase shift}. To make the arguments precise let us consider another non-radiating source, which is characterized by zero scalar potential and the following vector potential

\begin{equation}
\ve{A}_{cap}(t,\ve{r})=-i\omega t \sigma \ve{n}\iint_{D} d^2s\,\delta(\ve{r}-\ve{r}_s) \label{capacitor potential}
\end{equation} 

Notation here is the same as in formula \eqref{disk potential}. The electric field derived from \eqref{capacitor potential} is time-independent and coincides with the electric field of the disk-shaped dynamic anapole \eqref{disk electric field} taken at the moment $t=0$, $\ve{E}_{cap}(\ve{r})=\ve{E}(\ve{r},0)$. Since electric field is static and homogeneous, such a non-radiating source resembles an ordinary electric capacitor. However, by insisting that the fields outside the capacitor are strictly zero, we have also incorporated magnetic field at the boundary

\begin{eqnarray}
\ve{H}_{cap}(\ve{r},t)=-\omega t \sigma \int_{C} d\ve{l}\,\delta(\ve{r}-\ve{r}_l) \label{capacitor magnetic field}
\end{eqnarray} 

The phase shift for the contour penetrating the capacitor is given by magnetic flux, which linearly grows with time

\begin{eqnarray}
\delta\phi_{cap}(t)=-\omega t\f{e\sigma}{\hbar c} \label{capacitor phase shift}
\end{eqnarray}

Hereby, the oscillating anapole and non-radiating capacitor have the same electric field at $t=0$, but different magnetic fluxes leading to different phase shifts \eqref{time phase shift} and \eqref{capacitor phase shift}. 

Taking a slightly shifted perspective one can say that the essence of the AB effect is that the experimental setups equivalent in the classical sense do not in general exhibit the same behavior at the quantum level. Indeed, the standard AB experimental configuration with static magnetic field confined to an excluded region is classically equivalent to the complete absence of electromagnetic fields, yet it causes a measurable phase shift in probability amplitudes of a charged particle. Comparison of the oscillating anapole and non-radiating capacitor shows that a similar discrepancy is expected in the time-dependent experiment. Charged particles traveling fast enough would feel the same electric field in both cases (hence the classical equivalence) but would acquire different phase shifts resulting in observable discrepancies. Thus we conclude that in the dynamic case the main signature of the AB effect will still be present.

\section{Summary and discussion}
\new{We have shown that in the Weyl gauge the vector potential of an arbitrary NR source is spatially localized. We have also proven that this apparently local condition for an electromagnetic source to be non-radiating is equivalent (although in a non-trivial way) to the non-local criterion formulated in \cite{devaney1973radiating} by Devaney and Wolf. Using the obtained local non-radiating condition we confirmed that the relation $\ve{D}=-ik\ve{T}$, which holds for point NR sources, is also valid for spatially extended but physically small NR sources, where $\ve{D}$ and $\ve{T}$ are correspondingly the total electric and toroidal dipole moments of the system.}

\new{We have shown that any NR source can be viewed as a distribution of elementary dynamic anapoles -- NR point sources of the most fundamental type. Such an approach allows one to build concrete examples of spatially extended NR sources and study their properties. As an illustration, we considered a simple scenario for the dynamic version of the Aharonov-Bohm effect in the context of non-radiating sources. We came up with an explicit example of an NR source for which the phase shift in a dynamic experiment would arise exactly as in the static case, while retaining its dependence on time.}

Apart from the electrodynamics in general, the formalism developed here will be of particular importance for the fields of metamaterials and nanophotonics, which currently witness a surge of interest in the properties of the dynamic anapole and non-radiating systems (see \cite{raybould2015focused} and references therein). \new{Indeed, a number of recent works have already confirmed the key role of anapole excitations in controlling scattering properties of very simple electromagnetic systems, such as nanodisks and nanowires \cite{basharin2015dielectric, miroshnichenko2015nonradiating, liu2015toroidal, kim2015subwavelength, liu2015invisible, grinblat2016enhanced, xiang2016generic, tasolamprou2016toroidal, evlyukhin2016optical}. Correspondingly, one may want to revisit the analysis of the electromagnetic response of structurally more complex metamaterials, where the dynamic anapoles could underpin, for example, the microscopic mechanisms of electromagnetic transparency \cite{fedotov2013resonant} and high-Q effects. Our approach to non-radiating sources could be also useful in the analysis of non-radiating modes of antennas and scattering suppression in stealth applications. In particular, it might aid in designing stealth antennas and minimizing the radar cross-section of other elements that protrude from the airframe (such as meteorological sensors, guns, landing gear etc).}
\section*{Acknowledgments}

The authors gratefully acknowledge the financial support of the Ministry of Education and Science of the Russian Federation in the framework of Increase Competitiveness Program of NUST "MISiS" (\textnumero K4-2015-031) the Russian Foundation for Basic Research (Grant Agreement
No. 16-02-00789)

\appendix
\section{Criterion for NR sources according to Devaney and Wolf \label{nr criteria}}

In paper \cite{devaney1973radiating} the necessary and sufficient condition for a source to be non-radiating was established. It is formulated as follows. Expand harmonic current density $\ve{j}(\ve{r}, t)=e^{-i\omega t}\ve{j}(\ve{r})$ in terms of spatial Fourier modes $\ve{J}(\ve{p})$

\begin{eqnarray}
\ve{j}(\ve{r}, t)=e^{-i\omega t}\int d\ve{p} \,e^{i\ve{p}\ve{r}}\ve{J}(\ve{p})
\end{eqnarray}
and divide these into longitudinal and transverse components $\ve{J}(\ve{p})=\ve{J}_\perp(\ve{p})+\ve{J}_\parallel(\ve{p})$, where $\ve{J}_\perp(\ve{p})$ and $\ve{J}_\parallel(\ve{p})$ are orthogonal and parallel to vector $\ve{p}$ respectively. The current is non-radiating if and only if all the transverse components $\ve{J}_{\perp}(\ve{p})$ with $\ve{p}$, such that $|\ve{p}|=k=\omega/c$, are zero.

\new{We now prove that the above condition is equivalent to the vector potential being spatially localized in the gauge $\phi=0$. If the vector potential $\ve{A}(\ve{r})$ is localized, its Fourier components $\widetilde{\ve{A}}(\ve{p})$ are well-defined and simply related to the Fourier components of the current by the counterpart of equation \eqref{current via vector potential}, namely
\begin{eqnarray}
J_\al(\ve{p})=\f{c}{4\pi}\left((p^2-k^2)\delta_{\al\be}-p_\al p_\be\right)\tilde{A}_\be(\ve{p}) \label{j Fourier}
\end{eqnarray}
The transverse components of $\ve{J}(\ve{p})$ are proportional to $p^2-k^2$ and hence vanish at $|\ve{p}|=k$.}

\new{To prove the equivalence in backward direction we start with the standard formulas for the retarded potentials
\begin{eqnarray}
\phi(\ve{r})=\int d\ve{r}'\,\f{e^{ik|\ve{r}-\ve{r}'|}}{|\ve{r}-\ve{r}'|}\rho(\ve{r'})\\
\ve{A}(\ve{r})=\f1{c}\int d\ve{r}'\,\f{e^{ik|\ve{r}-\ve{r}'|}}{|\ve{r}-\ve{r}'|}\ve{j}(\ve{r'})
\end{eqnarray}
Making the gauge transformation which renders $\phi$ vanishing and using the continuity equation $-i\omega \rho+\operatorname{div}\ve{j}=0$ one arrives at the following expression for the vector potential
\begin{align}
\ve{A}(\ve{r})=\f1{k^2c}\int d\ve{r}'\,\f{e^{ik|\ve{r}-\ve{r}'|}}{|\ve{r}-\ve{r}'|}\left(k^2\ve{j}(\ve{r'})+\operatorname{grad}\operatorname{div}\ve{j}(\ve{r'})\right)
\end{align}
The latter features a convolution of co-ordinate functions, which can be re-written in terms of their Fourier components\footnote{Fourier representation of the Helmholtz equation Green's function  reads $\f{e^{ik|\ve{r}-\ve{r}'|}}{|\ve{r}-\ve{r}'|}=4\pi\int d\ve{p}\,\f{e^{i\ve{p}(\ve{r}-\ve{r}')}}{p^2-k^2}$.} 
\begin{align}
\ve{A}(\ve{r})=\f{4\pi}{k^2c}\int d\ve{p}\,\f{e^{i\ve{p}\ve{r}}}{p^2-k^2}\Big(k^2\ve{J}(\ve{p})-\ve{p}(\ve{p}\cdot\ve{J}(\ve{p}))\Big) \label{a via j}
\end{align}}
The integrand is an analytic function of $|\ve{p}|$ and if it decays fast enough at the complex infinity, the integral in $|\ve{p}|$ reduces to the residue at $|\ve{p}|=k$. It is easy to check that the integrand decays for sufficiently large $\ve{r}$. Indeed, since the current is spatially localized, its Fourier transform grows at most as $e^{-i\ve{p}\ve{r}_{max}}$ for some fixed $\ve{r}_{max}$, which defines the extent of the localization. For $|\ve{r}|>|\ve{r}_{max}|$ this growth is not enough to compensate for the decay of the factor $e^{i\ve{p}\ve{r}}$ in equation \eqref{a via j}. One therefore concludes that integral \eqref{a via j} only receives contributions from $\ve{p}$ such that $|\ve{p}|=k$. Note that for such $\ve{p}$ one has
\begin{multline}
k^2\ve{J}(\ve{p})-\ve{p}(\ve{p}\cdot\ve{J}(\ve{p}))=\\(k^2-\ve{p}^2)\ve{J}_\parallel(\ve{p})+k^2\ve{J}_\perp(\ve{p})=k^2\ve{J}_\perp(\ve{p})
\end{multline}
By assumption $J_\perp(\ve{p})=0$ at $|\ve{p}|=k$ and hence the integral (A6) vanishes for $|\ve{r}|>|\ve{r}_{max}|$ rendering $\ve{A}(\ve{r})$ as localized. 

We would like to point out that the localization of the vector potential can be proven with very little effort, directly from the definition of an NR source (see section \ref{general sources}). Our argument above basically yields another proof for the criterion of Devaney and Wolf, which, unlike the original work \cite{devaney1973radiating}, does not refer to the multipole expansion.

\section{Arbitrary potential from elementary dynamic anapoles\label{clarifications}}

Relation \eqref{potential from anapoles} implies that an arbitrary vector-potential field can be obtained as a superposition of the vector potentials of the elementary dynamic anapoles. The same result may be derived, perhaps with more comfort, by considering the chain of arguments in the reverse order. Indeed, any vector-potential field can be formally represented as a superposition of the delta functions \eqref{basis anapoles}. To understand what source produces a vector potential in the form of the delta-function one needs to substitute the latter into Maxwell's equations. This yields a distinct oscillating charge-current configuration, which corresponds to a combination of collocated electric and toroidal dipoles. By construction the vector potential (and hence electric and magnetic fields) of such a configuration is confined to a single point, which renders it as a point-like non-radiating source, i.e. the elementary dynamic anapole. 
 
One may detect a seeming contradiction here. More specifically, there is a strict relation between electric and toroidal dipoles in the DA, as defined by \eqref{d to tau relation}. Given that electromagnetic field of an arbitrary charge-current distribution can be expressed in terms of the fields of spatially distributed DAs, does this not imply that in fact any charge-current distribution has to satisfy \eqref{d to tau relation}? The caveat here is that two distinct charge-current distributions can produce the same electromagnetic fields (and potentials) if one of the distributions is not spatially localized. 
 
\begin{figure}[h]
\begin{center}
\begin{tikzpicture}
 
 \draw[dashed] (-4,0) circle (1);
 \filldraw[red] (-4,0) circle (.1);
 \node at (-4,-.4) {$r=0$};
 \node at (-4.3, 0) {$+$};
 \draw (-5.2,-1.2) rectangle (-2.8, 5);
 \node[below] at (-4,-1.2) {(a)};
 
 \node at (-2.5,0) {$\sim$};
 
 \draw[dashed] (-1,0) circle (1);
 \filldraw[red] (-1,0) circle (.1);
 \filldraw[blue] (-1,4) circle (.1);
 \node at (-1,-.4) {$r=0$};
 \node at (-1.3, 0) {$+$};
 \node at (-1,4.5) {$r=L\to \infty$};
 \node at (-1.3, 4) {$-$};
 \draw (-2.2,-1.2) rectangle (.2, 5);
 \node[below] at (-1,-1.2) {(b)};
 
 \node at (0.5,0) {$\sim$};
 
 \pgfmathsetmacro{\step}{0.6}
 \pgfmathsetmacro{\twosteps}{\step*2}
 
 \foreach \y in {0,\step,...,4}{
  \draw[line width=1] (2,\y) --++ (0,\step-0.1);
 
 }
 
 \foreach \y in {\step,\twosteps,...,4}{
  \filldraw[red] (2, \y) circle (.1);
  \node at (1.7,\y) {$+$};
  \filldraw[blue] (2, \y-.2) circle (.1); 
  \node at (1.7,\y-0.2) {$-$};
 }
 
 \draw[dashed] (2,0) circle (1);
 \filldraw[red] (2,0) circle (.1);
 \node at (1.7,0) {$+$};
 \filldraw[blue] (2,4) circle (.1);
 \node at (1.7,4) {$-$};

 \node at (2,-.4) {$r=0$};
 \node at (2,4.5) {$r=L\to\infty$};
 \draw (.8,-1.2) rectangle (3.2, 5);
 \node[below] at (2,-1.2) {(c)};
\end{tikzpicture}
\end{center}
\caption{Equivalence between fields of point charge and infinite dipole: (a) single charge, (b) large dipole, (c) large dipole as a chain of infinitesimal dipoles.}
\label{charge from dipoles}
\end{figure}
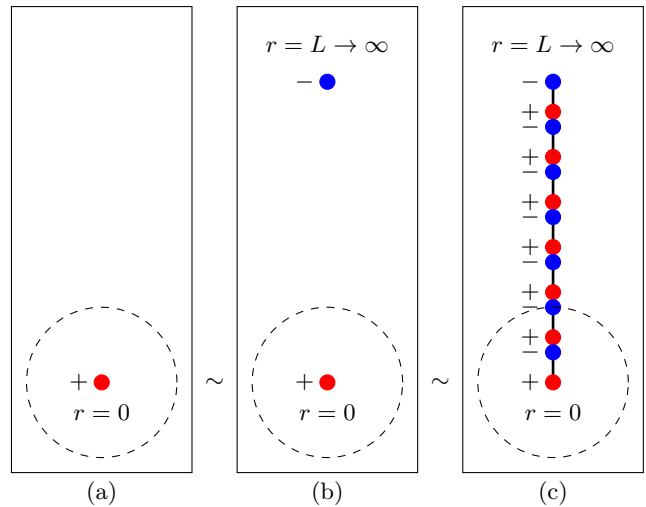

As the simplest example consider the electric field of an extended static dipole formed by two charges of opposite sign separated by distance $L$. The positive charge is placed at the origin, $r=0$, and the negative at $r=L$, see fig. \ref{charge from dipoles} (b). For large $L$ the influence of the negative charge in the vicinity of the positive charge (depicted by dashed circle) is negligible, and the electric field there is equivalent to the field of an isolated positive charge, fig. \ref{charge from dipoles} (a). In the formal limit of infinite $L$ the electric fields of the two charge configurations will also coincide in the rest of space. On the other hand, an 'infinite' dipole can be thought of as an assembly of infinitesimal dipoles arranged head-to-tail, fig. \ref{charge from dipoles} (c). Thus, an infinite number of point dipoles each carrying zero charge is able to precisely mimic the field of an isolated charged particle. 
 
The above considered example makes it clear that the relation \eqref{potential from anapoles} is rather formal. However, when the integration domain is \textit{finite}, which is exactly the case for non-radiating sources, \eqref{potential from anapoles} becomes much more useful and straightforward.

\bibliographystyle{apsrev4-1} 
\bibliography{bibfile}

\end{document}